\documentclass[superscriptaddress,secnumarabic,
amssymb,amsmath,nobibnotes,aps,prd,showkeys,showpacs,nofootinbib]{revtex4}%
\usepackage{graphicx}
\usepackage{float}
\usepackage{epsf}
\usepackage{bm}
\usepackage{amsmath}
\usepackage{amsfonts}
\usepackage{amssymb}
\usepackage{epstopdf}
\usepackage{natbib}
\usepackage{color}%
\providecommand{\U}[1]{\protect\rule{.1in}{.1in}}
\newcommand{\be}{\begin{equation}}
\newcommand{\ee}{\end{equation}}

\newcommand{\mincir}{\raise
-3.truept\hbox{\rlap{\hbox{$\sim$}}\raise4.truept\hbox{$<$}\ }}
\newcommand{\magcir}{\raise
-3.truept\hbox{\rlap{\hbox{$\sim$}}\raise4.truept\hbox{$>$}\ }}

\newtheorem{remark}{Remark}[section]
\usepackage{xcolor}

\newcommand{\R}{\mathbb{R}}
\begin{document}

\title{{ Relations between Newtonian and relativistic cosmology}}

\author{Jaume de Haro\footnote{E-mail: jaime.haro@upc.edu}}
\affiliation{Departament de Matem\`atiques, Universitat Polit\`ecnica de Catalunya, Diagonal 647, 08028 Barcelona, Spain}

\thispagestyle{empty}

\begin{abstract}


We start with 
the cosmic Friedmann equations, where we adopt a novel perspective rooted in a Lagrangian formulation grounded in Newtonian mechanics and the first law of thermodynamics. Our investigation operates under the assumption that the universe is populated by either a perfect fluid or a scalar field. By elucidating the intricate interplay between the Lagrangian formulation and the cosmic Friedmann equations, we uncover the fundamental principles governing the universe's dynamics within the framework of these elemental constituents.

In our concluding endeavor, we embark on the task of harmonizing the classical equations—namely, the conservation, Euler, and Poisson equations—with the principles of General Relativity. This undertaking seeks to extend these foundational equations to encompass the gravitational effects delineated by General Relativity, thus providing a comprehensive framework for understanding the behavior of matter and spacetime in the cosmic context.

\end{abstract}

\vspace{0.5cm}


\maketitle

\thispagestyle{empty}

\section{Introduction}

In the first segment of our study, we engage with the Friedmann equations, initially conceived by Alexander Friedmann in the 1920s through the lens of General Relativity's field equations. These equations furnish a comprehensive framework for comprehending the universe's evolutionary trajectory on cosmic scales, assuming homogeneity and isotropy while considering curvature and energy distribution.

It is noteworthy that Friedmann's groundbreaking contributions \cite{Friedmann}, depicting solutions encompassing both expanding and contracting universes, remained relatively obscure for a substantial duration \cite{Mostepanenko}. Einstein himself initially met Friedmann's cosmological findings with skepticism, contending that they deviated from the tenets of General Relativity. However, Einstein later retracted his critique, albeit without immediately embracing the notion of an expanding cosmos. The eventual recognition of the significance of Friedmann's work ensued with the discovery of the Hubble-Lemaître law, which bridged cosmology with fundamental physics. Through the amalgamation of General Relativity and thermodynamics, a comprehensive understanding of the universe's evolution materialized.

Proceeding along this trajectory, we explore an alternative avenue to deducing the Friedmann equations
{(also refer to  the early work by McCrea and Milne \cite{Milne}) and the one by Callan, Dicke and Peebles \cite{Peebles}}, leveraging Newtonian mechanics and the first law of thermodynamics. While these equations can be derived from Newton's formulations when conceiving the universe as a homogeneous dust fluid, employing the classical Lagrangian formalism furnishes a more exhaustive portrayal of dynamic systems in terms of kinetic and potential energy. Moreover, this formalism enables the contemplation of general perfect fluids and scalar fields, from which the Friedmann equations also emerge within the realm of Newtonian mechanics, thereby enriching our comprehension of the universe's evolutionary dynamics.

Concluding our investigation, we turn our attention to the generalization of the classical trio of equations - the continuity, Euler, and Poisson equations. The aim is to ensure their congruence with the principles of General Relativity, particularly under first-order perturbations. This endeavor necessitates extending these foundational equations to encompass the gravitational effects delineated by General Relativity, thus furnishing a holistic framework for apprehending the interplay between matter and spacetime in the presence of perturbations.

\section{Friedmann equations from General Relativity}

In this section, we will explore the Friedmann-Lemaître-Robertson-Walker (FLRW) metric (refer to \cite{Cook} for an interpretation of this metric):
\begin{eqnarray}\label{FLRW}
ds^2=-N^2(t)dt^2+a^2(t)\left(\frac{dr}{1-kr^2}+r^2(d\theta^2+\sin^2\theta d\phi^2)
\right),
\end{eqnarray}
where $N(t)$ denotes the lapse function, $k$ represents the spatial curvature, and $a(t)$ signifies the scale factor.

For this metric, the Ricci scalar is expressed as:
\begin{eqnarray}
R=6\left(\frac{1}{aN}\frac{d}{dt}\left(\frac{\dot{a}}{N} \right) +\frac{H^2}{N}+\frac{k}{a^2}\right),
\end{eqnarray}
where in this section, the "dot" signifies the derivative with respect to cosmic time, and $H=\dfrac{\dot{a}}{a}$ represents the Hubble rate.

Dealing with a homogeneous and isotropic universe filled with a perfect fluid (where pressure depends solely on energy density), the Einstein-Hilbert Lagrangian can be expressed as a function of the Ricci scalar, the space-time measure, and the matter content:
\begin{eqnarray}
L_{EH}=\frac{1}{2}R\sqrt{-g}-{8\pi G}\epsilon \sqrt{-g}=
\frac{1}{2}R a^3N-{8\pi G}\epsilon a^3N,
\end{eqnarray}
where $\epsilon$ represents the energy density, and $\sqrt{-g}=Na^3$ serves as the measure for the FLRW metric.

It is noteworthy that this Lagrangian can be reformulated as:
\begin{eqnarray}
L_{EH}=\frac{d}{dt}\left(\frac{\dot{a}a^2}{N}\right) -
\frac{3\dot{a}^2a}{N}+3kaN
-{8\pi G}\epsilon a^3N. \end{eqnarray}

Therefore, given that the first term is a total derivative, this Lagrangian is equivalent to:
\begin{eqnarray}\label{EHLagrangian}
\bar{L}_{EH}= -
\frac{3\dot{a}^2a}{N}+3kaN
-{8\pi G}\epsilon a^3N. \end{eqnarray}

Note that, 
the variation with respect the lapse leads to the so-called Hamiltonian constraint:
\begin{eqnarray}
   0= \dfrac{\partial \bar{L}_{EH}}{\partial N}=\dfrac{3\dot{a}^2a}{N^2}+3ka
    -{8\pi G}\epsilon a^3,\end{eqnarray}
where, after choosing  $N(t)=1$, one arrives to the first Friedmann equation:
\begin{eqnarray}\label{firstFriedmann}
    H^2+\dfrac{kc^2}{a^2}=\dfrac{8\pi G}{3}\epsilon.
\end{eqnarray}

To derive the dynamical equation, we perform a variation with respect to the scale factor. After setting $N=1$, the Euler-Lagrange equation
\begin{eqnarray}
\dfrac{d}{dt}\left(\dfrac{\partial \bar{L}_{EH}}{\partial \dot{a}}\right)
=
\frac{\partial \bar{L}_{EH}}{\partial a},
\end{eqnarray}
leads to 
\begin{eqnarray}
6\ddot{a}a+3\dot{a}^2={8\pi G} \frac{d}{da}(\epsilon a^3)-3ka,
\end{eqnarray}
where, due to the energy density's dependency solely on the scale factor, we replace $\frac{\partial}{\partial a}$ with $\frac{d}{da}$.

Next, assuming adiabatic evolution where the total entropy remains conserved, and utilizing the first law of thermodynamics
\begin{eqnarray}
d(\epsilon a^3)=-pda^3\Longrightarrow
\frac{d}{da}(\epsilon a^3)=-3p a^2,\end{eqnarray}
where $p$ represents pressure, we obtain:
\begin{eqnarray}
\frac{\ddot{a}}{a}=-{4\pi G} p-\frac{1}{2}\left(H^2+\frac{k}{a^2} \right).
\end{eqnarray}

From the first Friedmann equation (\ref{firstFriedmann}), we arrive at the acceleration, or second Friedmann equation:
\begin{eqnarray} \label{secondFriedmann}
\frac{\ddot{a}}{a}=-\frac{4\pi G}{3}( 3p+\epsilon).\end{eqnarray}

A final observation is warranted: In the FLRW space-time, the energy density and pressure of a perfect fluid solely depend on the scale factor, i.e., $\epsilon=\epsilon(a)$ and $p=p(a)$. Thus, we have:
\begin{eqnarray}\label{energydensity}
\epsilon(a)=\rho(a)-\frac{3}{a^3}\int^a \bar{a}^2p(\bar{a}) d\bar{a},
\end{eqnarray}
where $\rho$ represents mass density, assumed conserved, i.e., $d(a^3\rho)=0\Longrightarrow \rho=\frac{M}{a^3}$, with $M$ being the mass contained within volume $a^3$. Indeed, from the first law of Thermodynamics, $\frac{d\epsilon}{da}=-\frac{3(\epsilon+p)}{a}$, the solution of which is given by (\ref{energydensity}).

Consequently, the Einstein-Hilbert Lagrangian, as a function of the scale factor and lapse function, reads:
\begin{eqnarray}
\bar{L}_{EH}(a, \dot{a}, N)= -
\frac{3\dot{a}^2a}{N}+3kaN
-{8\pi G}\epsilon(a) a^3N, \end{eqnarray}
with $\epsilon(a)$ given by (\ref{energydensity}).

Finally, note that the energy density can be expressed as a function of mass density as follows \cite{Harko,Mendoza}:
\begin{eqnarray}
\epsilon(\rho)=\rho\left(1+\int^{\rho}\frac{p(\bar{\rho})}{\bar{\rho}^2}d\bar{\rho}
\right),
\end{eqnarray}
and thus, as a function of $\rho$ and $N$, the Einstein-Hilbert Lagrangian becomes:
\begin{eqnarray}
\bar{L}_{EH}(\rho, \dot{\rho}, N)= -
\frac{\dot{\rho}^2}{3N\rho^3}+3k
\left(\frac{M}{\rho} \right)^{1/3}N
-{8\pi G}\epsilon(\rho) \frac{MN}{\rho^3}. \end{eqnarray}

\subsection{Friedmann equations for an scalar field}

Scalar fields have proven to be highly advantageous in the study of cosmology. They have been instrumental in replicating the phenomenon of inflation \cite{Linde},  which explains the rapid expansion of the universe in its early stages. Additionally, scalar fields have also been utilized to model quintessence, a type of dark energy that is believed to be responsible for the accelerated expansion of the universe \cite{Tsujikawa}. These applications demonstrate the versatility and effectiveness of scalar fields in advancing our understanding of the cosmos.

Hence, in this section, we explore a homogeneous scalar field, denoted as $\phi$, minimally coupled with gravity, and derive the corresponding Friedmann equations within the framework of General Relativity. In this scenario, for the metric (\ref{FLRW}), the energy density and pressure are expressed as:
\begin{eqnarray}
\epsilon =\frac{\dot{\phi}^2}{2N^2}+V(\phi),\quad p =\frac{\dot{\phi}^2}{2N^2}-V(\phi),
\end{eqnarray}
and the corresponding Lagrangian is derived from (\ref{EHLagrangian}) by substituting the energy density with minus the pressure, yielding:
\begin{eqnarray}
\bar{L}_{EH}(a, \dot{a}, \phi,\dot{\phi}, N)= -
\frac{3\dot{a}^2a}{N}+3kaN
+{8\pi G} a^3
\left(\frac{\dot{\phi}^2}{2N}-NV(\phi)\right). \end{eqnarray}

Consequently, upon performing the variation with respect to the lapse function, we obtain:
\begin{eqnarray}
0= \frac{\partial \bar{L}_{EH}}{\partial N}=\frac{3\dot{a}^2a}{N^2}+3ka
-{8\pi G}
\left(\frac{\dot{\phi}^2}{2N^2}+V(\phi) \right) ,\end{eqnarray}
which, upon selecting $N=1$, transforms into the first Friedmann equation (\ref{firstFriedmann}).

On the other hand, when $N=1$, varying with respect to the scalar field yields the conservation equation:
\begin{eqnarray}\label{conservation}
\frac{d}{dt} \left(\frac{\partial \bar{L}_{EH}}{\partial \dot{\phi}}\right)=\frac{\partial \bar{L}_{EH}}{\partial {\phi}}
\Longrightarrow \ddot{\phi}+3H\dot{\phi}+\frac{\partial V}{\partial\phi}=0,
\end{eqnarray}
which equivalently represents the first law of thermodynamics.

Lastly, a straightforward computation demonstrates that the second Friedmann equation arises from the variation with respect to the scale factor.

\section{Friedmann equations from Newtonian mechanics}
We consider, in co-moving  coordinates,  a homogeneous large ball with a radius of $\bar{R}$ in Euclidean space (we can also consider $\bar{R}=+\infty$, but for finite radius, the total mass within the ball is finite $\bar{M}=\frac{4\pi}{3}\rho \bar{R}^3$, where $\rho$ is the mass density). Assume that the ball expands radially. This means that if $O$ is the center of the ball, a point $P$ within the ball at $t_0$ transforms into point $P_t$ at time $t$, and the distance from $O$ to $P_t$ is given by $d_{OP_t}\equiv d_{OP}(t)=a(t)d_{OP}=a(t)|\overrightarrow{OP}|$, where $a(t)$, with $a(t_0)=1$, is the scale factor. Furthermore, at time $t_0$, we consider the triangle $\widehat{POQ}$, which transforms into the equivalent triangle $\widehat{P_tOQ_t}$ at time $t$. Therefore, as $d_{OP_t}=a(t)d_{OP}$ and $d_{OQ_t}=a(t)d_{OQ}$, using Thales' theorem we find that for any points $P$ and $Q$ within the ball, $d_{P_tQ_t}=a(t)d_{PQ}$.

\

The relation shows that any ball, at $t_0$,  centered at a point $P$ with radius $R\ll \bar{R}$,  expands radially at the same rate as the original large ball. Additionally, the relative velocity between $P_t$ and $Q_t$ follows the Hubble-Lemaître law: \begin{eqnarray} \frac{d}{dt}(d_{P_tQ_t})= \dot{a}(t)d_{PQ} = H(t)d_{P_tQ_t}.\end{eqnarray}

\

The equation of motion for the scale factor in Newtonian mechanics is derived by considering 
a ball centered at a given point $P$ and initial radius $R$ at time $t_0$. At time $t$, the radial force at a given point $Q_t$ on the boundary of the ball is calculated to be
${\bf F}(Q_t)=f(a(t)R)
\frac{\overrightarrow{P_tQ_t}
}{a(t)R}$. To determine the function $f$, the flux entering the ball is computed as:
\begin{eqnarray}
    \Psi= \int {\bf F}.{\bf n}dS= 4\pi a^2(t)R^2f(a(t)R),
\end{eqnarray}
where ${\bf n}$ is the external normal to the sphere surrounding the ball and $dS$ is the measure of the sphere.
On the other hand, from the Poisson equation $\nabla.{\bf F}=-4\pi G \rho$,
and Gauss's theorem, the flux is also given by:
\begin{eqnarray}
    \Psi=\int \nabla. {\bf F}dV=-\frac{16 \pi^2G}{3}\rho a^3(t) R^3, 
\end{eqnarray}
which leads to the expression for $f$ as:
\begin{eqnarray}
f(a(t)R)=-\frac{4\pi G}{3}\rho a(t)R \Longrightarrow
{\bf F}(Q_t)=-\frac{4\pi G}{3}\rho \overrightarrow{P_tQ_t}.
\end{eqnarray}

Therefore, the acceleration experienced by a probe particle of mass $m$ at the point $Q$ due to the ball is determined by the second Newton's law as \cite{Mukhanov,Ryden,Guendelman}:
\begin{eqnarray}\label{Newton}
m\frac{d^2}{dt^2}({\overrightarrow{P_tQ_t}})= -\frac{4\pi G m}{3}\rho\overrightarrow{P_tQ_t}
    \Longrightarrow
    \ddot{a}=-\frac{4\pi G}{3}\rho a,    \end{eqnarray}
where we have used that 
$\overrightarrow{P_tQ_t}=a(t)\overrightarrow{PQ}$.

Here, it is important to recall that the constant $\kappa$ appearing in Einstein's field equations,
\begin{eqnarray}
R_{\mu\nu}-\frac{1}{2}g_{\mu\nu}R=\kappa T_{\mu\nu},
\end{eqnarray}
where $R_{\mu\nu}$ denotes the Ricci tensor, $g_{\mu\nu}$ the metric, and $T_{\mu\nu}$ the energy-stress tensor, is obtained under the assumption that the background is flat. Specifically, by approximating $g_{\mu\nu}=\eta_{\mu\nu}+h_{\mu\nu}$, where the background $\eta_{\mu\nu}$ is the Minkowski metric and $h_{\mu\nu}$ represents a small perturbation, Einstein's equations simplify to $R^0_0=\dfrac{\kappa}{2}T^0_0$, where
\begin{eqnarray}
R_0^0=-\frac{1}{2}\Delta g_{00},\quad\text{and}\quad T_0^0=\rho.
\end{eqnarray}

Considering that in the Newtonian approximation, $g_{00}$ and the Newtonian potential $\Phi$ are related by $g_{00}=-1-{2\Phi}$, and employing the Poisson equation $\Delta \Phi=4\pi G \rho$, we obtain:
\begin{eqnarray}
\Delta \Phi=\frac{\kappa \rho}{2}\Longrightarrow \kappa={8\pi G}.
\end{eqnarray}

Hence, it appears natural, as we have demonstrated, that the background is spatially flat and the volume of the ball is $\dfrac{4\pi}{3}a^3R^3$. However, as we shall see, the spatial curvature emerges in a natural manner.

Returning to Eq. (\ref{Newton}), we eliminate the mass $m$ to derive the second Friedmann equation for a dust field $(p=0)$:
\begin{eqnarray}
\frac{\ddot{a}}{a}=-\frac{4\pi G}{3}\rho.
\end{eqnarray}

This equation can be derived from the Lagrangian:
\begin{eqnarray}\label{NewtonianLagrangian}
L_N=
\frac{R^2\dot{a}^2}{2}+\frac{GM}{aR} = \frac{R^2\dot{a}^2}{2}+\frac{4\pi G}{3}R^2a^2\rho,
\end{eqnarray}
where $M=\frac{4\pi}{3}a^3R^3\rho$
represents the mass inside the ball.
Indeed, employing the Euler-Lagrange equation yields:
\begin{eqnarray}
\frac{d}{dt}\left(\frac{\partial L_N}{\partial \dot{a}}\right)=\frac{\partial L_N}{\partial {a}} \Longrightarrow\ddot{a}=\frac{4\pi G}{3}\frac{\partial}{\partial a}(a^2\rho)=-\frac{4\pi G}{3}\rho a,
\end{eqnarray}
where we have utilized mass conservation:
\begin{eqnarray}
\frac{\partial}{\partial a}(a^3\rho)=0
\Longrightarrow
\frac{\partial}{\partial a}(a^2\rho)=-a\rho.\end{eqnarray}

We observe that the radius $R$ of the chosen ball does not affect the dynamical equations. Thus, we set $R=1$.

To derive the second Friedmann equation for a general fluid field, we employ the relativistic equation $E=m$, relating the energy of a particle at rest to its mass. We substitute the mass density with the energy density in the Newtonian Lagrangian (\ref{NewtonianLagrangian}) with $R=1$, resulting in:
\begin{eqnarray}\label{NewtonianLagrangian1}
\bar{L}_N
=\frac{\dot{a}^2}{2}+\frac{4\pi G}{3}a^2\epsilon.
\end{eqnarray}

Using the Euler-Lagrange equation and the first law of thermodynamics,
\begin{eqnarray}
d(\epsilon a^3)=ad(\epsilon a^2)+\epsilon a^2da\Longrightarrow
-3p a^2da= ad(\epsilon a^2)+\epsilon a^2da
\Longrightarrow d(\epsilon a^2)=-(3p+\epsilon)ada,
\end{eqnarray}
we readily derive (\ref{secondFriedmann}).

The next step is to obtain the first Friedmann equation. This can be achieved by combining the second equation with the first law of Thermodynamics, expressed as follows:
\begin{eqnarray}
\dot{\epsilon}=-3H(\epsilon+p).
\end{eqnarray}

Firstly, we rewrite (\ref{secondFriedmann}) as:
\begin{eqnarray}\label{A}
\frac{\ddot{a}}{a}=-{4\pi G}( p+\epsilon)+\frac{8\pi G}{3}\epsilon.\end{eqnarray}
Then, we calculate
\begin{eqnarray}
\frac{d H^2}{dt}=2H\frac{\ddot{a}}{a}-2H^3.
\end{eqnarray}
Inserting (\ref{A}) into it, we obtain:
\begin{eqnarray}
\frac{d}{dt}\left(H^2-
\frac{8\pi G}{3}\epsilon
\right)=-2H\left(H^2-
\frac{8\pi G}{3}\epsilon
\right)\Longrightarrow
\frac{d\left(H^2-
\frac{8\pi G}{3}\epsilon
\right)}{H^2-
\frac{8\pi G}{3}\epsilon}=-2\frac{da}{a}, \end{eqnarray}
whose solution is given by
\begin{eqnarray}
H^2-\frac{8\pi G}{3}\epsilon=\frac{C}{a^2},\end{eqnarray}
and by setting the constant of integration $C$ equal to $-k$, we obtain the first Friedmann equation.

Hence, the Newtonian Lagrangian in terms of the scale factor is given by:
\begin{eqnarray}\label{NewtonianLagrangian2}
\bar{L}_N(a,\dot{a})
=\frac{\dot{a}^2}{2}+\frac{4\pi G}{3}a^2\epsilon(a),
\end{eqnarray}
with $\epsilon(a)$ given by (\ref{energydensity}).

Finally, considering that the energy of a homogeneous ball of radius $a$ is $E=\dfrac{4\pi}{3}a^3\epsilon$, the Newtonian Lagrangian appears as $\bar{L}_N=E_{\text{kin}}-V$, where $E_{\text{kin}}=\dfrac{\dot{a}^2}{2}$ is the kinetic energy per unit mass and $V=-\dfrac{GE}{a}$ is the gravitational potential generated by the ball with rest mass $E$.

\subsection{Friedmann equations for an scalar field}

In a manner analogous to relativistic cosmology, when dealing with a scalar field, we replace $\epsilon$ with   $-p$ in the Newtonian Lagrangian (\ref{NewtonianLagrangian2}).

Let $\phi'$ denote the derivative of the scalar field with respect to the scale factor. We have $\dot{\phi}=\dot{a}\phi'$, and thus, the pressure takes the form
$p=\frac{\dot{a}^2(\phi')^2}{2}-V(\phi)$,
resulting in the Lagrangian:
\begin{eqnarray}\label{NewtonianLagrangian3}
    \bar{L}_N(a,\dot{a})
    =\frac{\dot{a}^2}{2}-\frac{4\pi G}{3}a^2 \left(\frac{\dot{a}^2(\phi')^2}{2}-V(\phi)\right),
    \end{eqnarray}
where the scalar field is now a function of the scale factor.

Firstly, the first law of thermodynamics, expressed as $\dfrac{d}{dt}(\epsilon a^3)=-3pa^2 \dot{a}$, yields the conservation equation (\ref{conservation}). Then, upon variation with respect to the scale factor, we obtain the second Friedmann equation (\ref{secondFriedmann}). Essentially, this yields:

\begin{eqnarray}
    \frac{\partial \bar{L}_N}{\partial \dot{a}}=
    \dot{a}-\frac{4\pi G}{3}a^2 \dot{a}(\phi')^2=\dot{a}-\frac{4\pi G}{3}\frac{a^2} {\dot{a}}\dot{\phi}^2,
    \end{eqnarray}
and thus, 
\begin{eqnarray}
    \frac{d}{dt}\left(\frac{\partial \bar{L}_N}{\partial \dot{a}}\right)= \ddot{a}-\frac{4\pi G}{3}\frac{d}{dt}\left(\frac{a^2} {\dot{a}}\dot{\phi}^2\right)
    =\ddot{a}-\frac{4\pi G}{3}
    \left(a^2\dot{\phi}\frac{d}{dt}
    \left(\frac{\dot{\phi}}{\dot{a}}\right)+2a\dot{\phi}^2+\frac{a}{H}\dot{\phi}\ddot{\phi} 
    \right).\end{eqnarray}

On the other hand, 
\begin{eqnarray}
\frac{\partial \bar{L}_N}{\partial {a}}=-\frac{4\pi G}{3}\left( \dot{a}^2a(\phi')^2+\dot{a}^2a^2\phi'\phi''-a^2\frac{\partial V}{\partial \phi}\phi'-2aV\right)
=-\frac{4\pi G}{3}\left( a\dot{\phi}^2+a^2\dot{\phi}\frac{d}{dt}
    \left(\frac{\dot{\phi}}{\dot{a}}\right)-
\frac{a^2}{\dot{a}}\frac{\partial V}{\partial \phi}\dot{\phi}-2aV\right).
\end{eqnarray}

Then, applying the Euler-Lagrange equation, we arrive at:
\begin{eqnarray}
\ddot{a}=-\frac{4\pi G}{3}\left( -a(\dot{\phi}^2-2V)
-\frac{a}{H}\dot{\phi}\left(\frac{\partial V}{\partial \phi}+\ddot{\phi} \right)
\right).
\end{eqnarray}
Utilizing the conservation equation (\ref{conservation}), we derive the second Friedmann equation as:
\begin{eqnarray}
\ddot{a}=-\frac{8\pi G}{3}(\dot{\phi}^2-V)a=-\frac{4\pi G}{3}(\epsilon+3p)a, \end{eqnarray}
where we have employed the relation $\epsilon+3p=2(\dot{\phi}^2-V)$.

\subsection{Application to open systems}

The Newtonian formulation extends to open systems, such as adiabatic systems where matter creation is permitted, conserving the total entropy. The first law of thermodynamics in such a scenario reads \cite{Prigogine,Prigogine1}:
\begin{eqnarray}
d(a^3\epsilon)=-pda^3+\frac{(\epsilon+p)a^3}{\bar{N}}d\bar{N},
\end{eqnarray}
where $\bar{N}(t)$ represents the number of produced particles at time $t$.
Equivalently, this equation can be expressed as
\begin{eqnarray}
\dot{\epsilon}+3H\left(1-\frac{\dot{\bar{N}}}{3H\bar{N}} \right)(\epsilon+p)=0.
\end{eqnarray}

Employing the Lagrangian $\bar{L}_N$, the second Friedmann equation becomes:
\begin{eqnarray}
\frac{\ddot{a}}{a}=-\frac{4\pi G}{3}( 3p+\epsilon)+\frac{4\pi G}{3}\frac{\dot{\bar{N}}}{H\bar{N}}(p+\epsilon)
\Longleftrightarrow
\frac{\ddot{a}}{a}=-{4\pi G}\left(1-\frac{\dot{\bar{N}}}{3H\bar{N}} \right)(p+\epsilon)+\frac{8\pi G}{3}\epsilon,
\end{eqnarray}
while the first one remains (\ref{firstFriedmann}).

Moreover, combining both Friedmann equations provides insight into the evolution of the Hubble rate. Specifically, considering $\dot{H}= \dfrac{\ddot{a}}{a}-H^2$, we derive:
\begin{eqnarray}\label{dotH}
\dot{H}= -{4\pi G}\left(1-\frac{\dot{\bar{N}}}{3H\bar{N}} \right)(p+\epsilon)+\frac{k}{a^2}.\end{eqnarray}

This equation admits analytical solutions for linear Equations of State ($p=w \epsilon$, with $w$ constant), particularly in spatially flat scenarios, across several open models.

For various functions $\Gamma$ defining the particle production rate $\Gamma=\dfrac{\dot{\bar{N}}}{\bar{N}}$, one can analytically determine the universe's evolution \cite{Haro,Haro1}. For instance, in the case of a constant $\Gamma>0$, the solution derived from the first Friedmann equation and (\ref{dotH}) yields:
\begin{eqnarray}
\dot{H}= -{4\pi G}\left(1-\frac{\Gamma}{3H} \right)(1+w)\epsilon=-\frac{3(1+w)}{2}\left(1-\frac{\Gamma}{3H} \right)H^2, \end{eqnarray}
with the solution:
\begin{eqnarray}
H(t)=\frac{\Gamma}{3}\exp\left(
\frac{\Gamma(1+w)}{2}(t-t_s)\right)\left(
\exp\left(
\frac{\Gamma(1+w)}{2}(t-t_s)\right)-1 \right)^{-1}.
\end{eqnarray}

This solution indicates a big bang singularity at $t=t_s$ ($H(t_s)=+\infty$), transitioning to a de Sitter phase at late times, where $H(t)\cong \Gamma/3$.

In closing, it is worth noting that more generalized particle production rates, such as $\Gamma(H)=-\Gamma_0+mH+n/H$, where $\Gamma_0$, $m$, and $n$ are constants, have been extensively investigated \cite{Haro1,Pan}. These models predict early and late accelerated expansion phases for various parameter values.

\section{Perturbations in classical mechanics:  perfect fluids}

This section endeavors to generalize the fundamental classical equations in fluid dynamics to align, at least to the first order of perturbations, with Einstein's field equations.

Expanding classical fluid dynamics to incorporate the principles of General Relativity marks a significant step in our understanding of the universe's behavior. By extending classical equations, we aim to capture the intricate dynamics of spacetime curvature influenced by fluid distributions.

In this pursuit, it becomes imperative to reconcile the robust framework of classical fluid dynamics with the profound insights offered by General Relativity. Achieving this alignment facilitates a deeper comprehension of how matter and energy interact with the fabric of spacetime.

\subsection{First law of thermodynamics}

Let $\varphi_t: \mathbb{R}^3 \rightarrow \mathbb{R}^3$ be the flow of a perfect fluid, with $\varphi_0 = \text{Id}$. We define the vector velocity ${\bf v}(\varphi_t({\bf q}), t) = \dfrac{d\varphi_t({\bf q})}{dt}$.

Then, we arrive at the crucial result, as outlined in \cite{Girbau}:
\begin{eqnarray}
\left[\frac{d}{dt}\int_{\varphi_t(V)} f({\bf q}, t) dV\right]_{t=\bar{t}} =
\int_{\varphi_{\bar{t}}(V)}\left( \frac{\partial f}{\partial t}+ \nabla_{\bf q} \cdot (f{\bf v}) \right)_{t=\bar{t}}dV,
\end{eqnarray}
where $\nabla{\bf q} \cdot {\bf u}$ denotes the divergence of the vector field ${\bf u}$.

\

Applying this result to the first law of thermodynamics:
\begin{eqnarray}
\left[\frac{d}{dt}\int_{\varphi_t(V)} \epsilon({\bf q}, t) dV\right]_{t=\bar{t}} =
-p({\bf q}, t)
\left[\frac{d}{dt}\int_{\varphi_t(V)} 1 dV\right]_{t=\bar{t}}. \end{eqnarray}
Here, once again, $\epsilon$ denotes the energy density of the fluid and $p$ its pressure. This yields the conservation equation:
\begin{eqnarray}
\frac{\partial \epsilon}{\partial t}+\nabla \cdot (\epsilon {\bf v}) = -p\nabla \cdot {\bf v}.
\end{eqnarray}

Next, we consider an expanding universe described by the flat FLRW metric $ds^2=-dt^2+a^2d{\bf q}^2$. The element of volume is given by $dV=a^3dq_1dq_2dq_3$, and in differential form, the first law of thermodynamics becomes:
\begin{eqnarray}
\dot{\epsilon}+3H(\epsilon+p)+
\nabla_{\bf q}\cdot(\epsilon{\bf v})
+p\nabla_{\bf q}\cdot{\bf v}=0,\end{eqnarray}
which, up to first order ($\epsilon=\epsilon_0+\delta\epsilon$ and $p=p_0+\delta p$), leads to:
\begin{eqnarray}
\dot{\epsilon}_0+3H(\epsilon_0+p_0)=0\qquad \text{and}\qquad
\dot{\delta\epsilon}
+(\epsilon_0+p_0)\nabla{\bf q}\cdot{\bf v}=0.\end{eqnarray}

We can also introduce gravity by considering the following metric in the weak field approximation $|\Phi|\ll 1$:
\begin{eqnarray}
d\tau^2=(1+2\Phi({\bf q}))dt^2- { (1-2\Phi({\bf q})) }d{\bf q}^2,
\end{eqnarray}
which coincides with formula (106.3) of \cite{Landau} (also obtained in Einstein's book "The Meaning of Relativity" \cite{A.Einstein}). In modern language, this is referred to as the "Newtonian gauge".

Then, applying the first law of thermodynamics to this metric including the expansion of the universe, i.e., to
\begin{eqnarray}
d\tau^2=(1+2\Phi({\bf q}, t))dt^2- a^2(t){ (1-2\Phi({\bf q},t)) }d{\bf q}^2,
\end{eqnarray}
one obtains the first-order perturbed equation
\begin{eqnarray}
\dot{\delta \epsilon}+{(\epsilon_0+p_0)}\nabla_{\bf q}\cdot{\bf u}+3{ H}(\delta\epsilon+\delta p)-3(\epsilon_0+p_0)\dot{\Phi}=0.
\end{eqnarray}

At this point, it is useful to use the notation
${\bf u}\equiv\dfrac{d\bf q}{d\eta}= a{\bf v}$ being $\eta$ the conformal time  and, $\nabla\equiv \dfrac{1}{a}\nabla_{{\bf q}}$, obtaining:
\begin{eqnarray}
\dot{\delta \epsilon}+{(\epsilon_0+p_0)}\nabla\cdot{\bf u}+3{ H}(\delta\epsilon+\delta p)-3(\epsilon_0+p_0)\dot{\Phi}=0.
\end{eqnarray}

\begin{remark}
It is important to recall that this equation is the same as the linearized equation $\nabla_{\mu}T^{\mu 0}=0$, where
\begin{eqnarray}
T^{\mu\nu}=(\epsilon+p)u^{\mu}u^{\nu}+pg^{\mu\nu}
\end{eqnarray}
is the stress-energy tensor.
\end{remark}

\subsection{Euler's equation}

First of all, we recall that for a dust fluid, i.e., $|p|\ll \epsilon\cong \rho$, and the element of volume $dV=dq_1dq_2dq_3$, the classical Euler's equation can be written as:
\begin{eqnarray}\label{Eulerintegral}
\left[\frac{d}{dt}\int_{\varphi_t(V)} 
\rho{\bf v} dV\right]_{t=\bar{t}}=
\int_{\partial \varphi_{\bar t}(V)} {\bf \mathcal T}({\bf n})dS
-\int_{\varphi_{\bar t}(V)}\rho\nabla \Phi
dV,\end{eqnarray}
where $ {\bf \mathcal T}:\R^3\longrightarrow \R^3$ is the stress tensor, $dS$ is the element of area, and ${\bf n}$ is the external unit vector to the boundary. Taking into account that for a perfect fluid one has
$ {\bf \mathcal T}({\bf n})=-p{\bf n}$ and, from the Gauss theorem, the Euler equation in the differential form becomes:
\begin{eqnarray}\label{Euler}
\partial_t(\rho{\bf v})+ \nabla_{{\bf q}}.(\rho{\bf v}){\bf v}+
\rho{\bf v}.\nabla_{{\bf q}}{\bf v}+\nabla_{{\bf q}} p+\rho \nabla_{{\bf q}} \Phi=0 \qquad \mbox{ or}
\qquad
\dot{{\bf v}}+
{\bf v}.\nabla_{{\bf q}}{\bf v}+\frac{1}{\rho}\nabla_{{\bf q}} p+ \nabla_{{\bf q}} \Phi=0, \end{eqnarray}
where we have used the first law of thermodynamics for a dust fluid or the continuity equation $\dot{\rho}+\nabla_{{\bf q}}.(\rho{\bf v})=0$.

Note that the equation (\ref{Euler}) is incompatible with special relativity. For this reason, we have to compare it with the conservation law $\partial_{\mu} T^{\mu k}=0$ in the Minkowski spacetime:
\begin{eqnarray}
\ \partial_t((\epsilon+p){\bf v})+ \nabla{\bf q}.((\epsilon+p){\bf v}){\bf v}+
(\epsilon+p){\bf v}.\nabla_{\bf q} {\bf v}+\nabla_{\bf q} p=0. \end{eqnarray}

Therefore, the Euler equation in an expanding universe compatible with special relativity is obtained by replacing the mass density $\rho$ by the heat function per unit volume $(\epsilon+p)$ \cite{fluid}, the velocity ${\bf v}$ by
$a^2{\bf v}$
and using the element of volume $dV=a^3dq_1dq_2dq_3$
in (\ref{Eulerintegral}):
\begin{eqnarray}\label{Eulerintegral}
\left[\frac{d}{dt}\int_{\varphi_t(V)} a^2(\epsilon+p){\bf v} dV\right]_{t=\bar{t}}=
\int_{\partial \varphi_{\bar t}(V)} {\bf \mathcal T}({\bf n})dS
-\int_{\varphi_{\bar t}(V)}(\epsilon+p)\nabla_{\bf q} \Phi
dV,\end{eqnarray}
obtaining, using once again the notation
${\bf u}=a{\bf v}$ and, $\nabla =\dfrac{1}{a}\nabla_{\bf q}$,
at the first order of perturbations:
{ \begin{eqnarray}
\partial_t((\epsilon_0+p_0){\bf u})+4H (\epsilon_0+p_0){\bf u} +\nabla \delta p+(\epsilon_0+p_0)\nabla \Phi=0, \end{eqnarray}}
which is equivalent, up to linear terms, to the equation $\nabla_{\mu}T^{\mu k}=0$.

\subsection{Poisson's equation}

We start with the classical Poisson equation:
\begin{eqnarray}
\Delta_{\bf q}\Phi=4\pi G a^2 \rho \Longrightarrow\Delta\Phi=4\pi G \rho,
\end{eqnarray}
where, once again, $\rho$ denotes the mass density.

The last equation is the Hamiltonian constraint \cite{Eric}:
\begin{eqnarray}
\mathcal{I}+ \mathcal{R}=16\pi G E,
\end{eqnarray}
where $\mathcal{R}$ is the intrinsic curvature (the spatial curvature),
$
{\mathcal I}=K^2-K_{ij}K^{ij}$, where
$
K_{ij}=\frac{1}{\sqrt{N}}g(\nabla_{\partial_i}\partial_j, \partial_t)
$,
is the extrinsic curvature and $E=T_{00}n^0n^0$, with $n^{\alpha}=\dfrac{1}{\sqrt{1+2\Phi}}\partial_t$ as the unit time vector.

In our case, we have:
\begin{eqnarray}
{\mathcal I}\cong {6}(1-2\Phi)\left(
H-\dot{\Phi}
\right)^2,
\qquad
{\mathcal R}=4\Delta \Phi
\qquad\mbox{and}\qquad
E\cong \epsilon.
\end{eqnarray}

Therefore, the third equation is:
\begin{eqnarray}\label{hamiltonian}
2 \Delta \Phi+
{3}\left(
H^2-2H^2\Phi-2H{\dot{\Phi}}
\right)
=8\pi G\epsilon,
\end{eqnarray}
which leads to the perturbed equation:

\begin{eqnarray}
\Delta \Phi
-{3}\left(
H^2\Phi+H{\dot\Phi}
\right)
=4\pi G\delta\epsilon. \end{eqnarray}

\begin{remark}
One could understand $\frac{1}{16\pi G}{\mathcal I}$ as the kinetic energy of the field and $\frac{1}{16\pi G}{\mathcal R}
$
as its potential energy.
\end{remark}

\

{Finally, note that when one introduces the new variable $\psi=a(1-2\Phi)$, 
the last equation becomes a heat equation  with a heat source:
\begin{eqnarray}
    3H\dot{\psi}-\Delta\psi=
    {8\pi G a}\epsilon,
\end{eqnarray}
which can be seen as the combination of the first Friedmann equation and the Poisson one, that is, the combination of:
\begin{eqnarray}
    3H\dot{a}=8\pi G a\epsilon_0,\qquad
    \Delta \Phi=4\pi G\rho.
\end{eqnarray}

Note also that, in Fourier space, i.e.,
writing $\psi=\int_{\R^3}\psi_{\bf k}e^{i{\bf q}.{\bf k}}d{\bf k}^3$, 
its solution is:
\begin{eqnarray}
    \psi_{\bf k}(a)=\mathcal{K}_{\bf k}(a;a_i)\psi_{\bf k}(a_i)+\int_{a_i}^a 
    \mathcal{K}_{\bf k}(a;\bar{a})    \delta_{\bf k}(\bar{a})d\bar{a},
\end{eqnarray}
where we have introduced the notation
$\delta_{\bf k}=\frac{\epsilon_k}{\epsilon_0}$ and 
\begin{eqnarray}
    \mathcal{K}_{\bf k}(a;b)=\exp\left(-\frac{|{\bf k}|^2}{8\pi G}\int_b^a \frac{dz}{z^3\epsilon_0(z)}\right),
    \end{eqnarray}
is the kernel of the homogeneous heat equation.

}

\subsection{Generalization of the three classical equations}

The generalization of the three "classical" equations,
for the volume element $ dV=a^3(1-3\Phi)dq_1dq_2dq_3$,
which,  only up to linear order,  are equivalent to the Einstein's field equations,  are:
\begin{eqnarray}\left\{\begin{array}{ccc}
    \left[\frac{d}{dt}\int_{\varphi_t(V)} \epsilon dV\right]_{t=\bar{t}} &=&
    -p
    \left[\frac{d}{dt}\int_{\varphi_t(V)} 1 dV\right]_{t=\bar{t}} \\ 
& & \\
    \left[\frac{d}{dt}\int_{\varphi_t(V)} a(\epsilon+p){\bf u} dV\right]_{t=\bar{t}}&=&
    -\int_{\partial \varphi_{\bar t}(V)} p{\bf n}dS
    -\int_{\varphi_{\bar t}(V)}a(\epsilon+p)\nabla\Phi
    dV\\ & & \\
   2 \Delta \Phi
+
{3}\left(
    H^2-2H^2\Phi-2H{\dot{\Phi}}
    \right)
    &= &8\pi G\epsilon.
    \end{array}\right.
    \end{eqnarray}

In differential form,  these "classical" equations can be approximated by:
\begin{eqnarray}
\label{newtonequations}
\left\{ \begin{array}{ccc}
D_t{\epsilon}+(\epsilon+p)\left[3(H-\dot{\Phi})
+\nabla.  {\bf u}\right]   &= &0 \\
&&\\
D_t \left( {(\epsilon+p)}{\bf u}    \right)+
(\epsilon+p)\left[4H{\bf u}
+ (\nabla.{\bf u}){\bf u} +\nabla\Phi\right]
+\nabla {p}   
&=&0 \\ &&\\
   2 \Delta \Phi
+
{3}\left(
    H^2-2H^2\Phi-2H{\dot{\Phi}}
    \right)
    &= &8\pi G\epsilon, 
  \end{array} \right.
\end{eqnarray}
where we have introduced the standard notation in fluid mechanics for the total time-derivative:  $D_t f=\partial_t f+{\bf u}.\nabla f$.

\

We can {\color{red}also} observe that the first and second equations, i.e., the first law of thermodynamics and the generalization of the Euler's equation, up to linear order, correspond to the conservation of the energy-stress tensor: 
$\nabla_{\mu} T^{\mu\nu}=0$.
And the last one is the generalization of the Poisson equation.

In a static universe, for a dust fluid and a weak static potential, we recover the classical equations, namely, the continuity, Euler, and Poisson equations:
\begin{eqnarray}\left\{ \begin{array}{ccc}
D_t{\rho}+\rho\nabla.  {\bf u}  &= &0 \\ &&\\
D_t {\bf u}    +\frac{1}{\rho}\nabla {p}   
+
\nabla\Phi
&=&0 \\ &&\\
   \Delta \Phi
    &= &4\pi G\rho. \end{array} \right.
    \end{eqnarray}
 
 \
 
 And the linear order perturbed equations are the same as in General Relativity:
 \begin{eqnarray}\left\{\begin{array}{ccc}
    \dot{\delta\epsilon}
    + (\epsilon_0+p_0)[\nabla. 
    {\bf u}-3\dot{\Phi}]    +3H(\delta\epsilon+\delta p)&=&0
\\  &&\\
\partial_t\left(
    (\epsilon_0+p_0){\bf u}
    \right)+(\epsilon_0+p_0)[4H{\bf u}
     +
    \nabla \Phi]
     + \nabla \delta p      & = & 0
    \\&&\\
    \Delta \Phi
-{3}\left(
    H^2\Phi+H{\dot{\Phi}}
    \right)
    &= & 4\pi G\delta\epsilon. \end{array}\right.
     \end{eqnarray}

\

To conclude this section, it is important to recognize that the last equation in (\ref{newtonequations}) serves as a constraint, specifically the Hamiltonian constraint. However, by combining all three equations, we can derive the following dynamical equation for the Newtonian potential:
\begin{eqnarray}
\frac{6}{a^5}\partial_t{(a^5\dot{\Phi})}-
    2\Delta\Phi-{R_0}(1-2\Phi)={8\pi G}T\Longleftrightarrow
6\ddot{\Phi}-2\Delta\Phi+30H\dot{\Phi}-{R_0}(1-2\Phi)={8\pi G}T,\end{eqnarray}
where $T=3p-\epsilon$ is the trace of the stress-energy tensor and $R_0=6(\dot{H}+2H^2)$ the zero order Ricci scalar.

\

This equation encompasses both the second Friedmann equation and the Poisson equation. Hence, the three dynamical equations in Newtonian theory, which encapsulate the Friedmann equations and the perturbed equations of General Relativity, are:
\begin{eqnarray}
\label{newtonequations1}
\left\{ \begin{array}{ccc}
D_t{\epsilon}+(\epsilon+p)\left[3(H-\dot{\Phi})
+\nabla.  {\bf u}\right]   &= &0 \\&&\\
D_t \left( {(\epsilon+p)}{\bf u}    \right)+
(\epsilon+p)\left[4H{\bf u}
+ (\nabla.{\bf u}){\bf u} +\nabla\Phi\right]
+\nabla {p}   
&=&0 \\&&\\
3\ddot{\Phi}-\Delta\Phi+15H\dot{\Phi}-3(\dot{H}+2H^2)(1-2\Phi)&=&{4\pi G}(3p-\epsilon)
. \end{array} \right.
    \end{eqnarray}

\

A final remark is in order: By utilizing the conformal time $d\eta=\frac{dt}{a}$ and introducing the new variable $\bar{\bf q}=\sqrt{3}{\bf q}$, the dynamical equation for the Newtonian potential takes the form:

\begin{eqnarray}\label{conformal}
3\left[ (a^4\Phi')'-a^4\Delta_{\bar{\bf q}}\Phi\right]+a^6R_0\Phi=4\pi Ga^6\delta T,
\end{eqnarray}

where $\delta T=3\delta p-\delta \epsilon$ represents the linear perturbation of the stress-energy tensor.

We can easily see that this equation can be obtained from the variation of the Lagrangian with respect to the Newtonian potential:

\begin{eqnarray}
\mathcal{L}=a^4\left[
\frac{(\Phi')^2}{2}-\frac{|\nabla_{\bar{\bf q}}\Phi|^2}{2}-\frac{a^2R_0}{6}\Phi^2+\frac{4\pi Ga^2}{3}\delta T \Phi
\right],
\end{eqnarray}

and we can recognize its similarity with the Lagrangian corresponding to a massless scalar field, $\phi$, conformally coupled with gravity:

\begin{eqnarray}
\mathcal{L}=\frac{a^2}{2}\left[
{(\phi')^2}-{|\nabla_{{\bf q}}\phi|^2}-\frac{a^2R_0}{6}\phi^2
\right].
\end{eqnarray}

\

On the other hand, performing the transformation $\Phi=\dfrac{\bar\Phi}{a^2}$, the dynamical equation (\ref{conformal}) becomes:
\begin{eqnarray}
    \bar\Phi''-\Delta_{\bar{\bf q}}\bar\Phi
    -2{\mathcal H}^2\bar\Phi=\dfrac{4\pi G a^4}{3}\delta T,
    \end{eqnarray}
which can be obtained from the variation of the Lagrangian 
\begin{eqnarray}
{\mathcal L}=\dfrac{1}{2}
    \left[{(\bar\Phi')^2}-{|\nabla_{\bar{\bf q}}\bar\Phi|^2}+2{\mathcal H}^2\bar\Phi^2\right]+\dfrac{4\pi Ga^4}{3}\delta T \bar\Phi,
\end{eqnarray}
where the firsts three terms
resemble those of a harmonic oscillator with a time-dependent frequency $\sqrt{2}{\mathcal H}$,  and  the last one represents the coupling between the potential and the stress-energy  tensor. Additionally, in a static universe, the Newtonian potential satisfies the typical wave equation under the action of a mass source:
\begin{eqnarray}
    \ddot{\Phi}-\Delta_{\bar{\bf x}}\Phi
    =\dfrac{4\pi G}{3}\delta T,
    \end{eqnarray}
where $\bar{\bf x}=a\bar{\bf q}$.

\

Finally, we recast our equations in conformal time, coordinates {${{\bf q}}$, velocity ${\bf u}=\dfrac{d{{\bf q}}}{d\eta}$,  and potential $\bar\Phi$, i.e., using the metric
$ds^2=-(a^2+2\bar\Phi)d\eta^2+
(a^2-2\bar\Phi)d{{\bf q}}^2$:
\begin{eqnarray}
\label{newtonequations2}
\left\{ \begin{array}{ccc}
 D_{\eta}{\epsilon}+(\epsilon+p)\left[3\left({\mathcal H}-{\dfrac{\bar\Phi'}{a^2}}+2{\mathcal H}{\dfrac{\bar\Phi}{a^2}}\right)
+\nabla_{ {\bf q}}.  {\bf u}\right]   &= &0 \\&&\\
D_{\eta} \left( {(\epsilon+p)}{{\bf u}}    \right)+
(\epsilon+p)\left[4{\mathcal H}{\bf u}
+ (\nabla_{{\bf q}}.{\bf u}){\bf u} +\dfrac{1}{a^2}\nabla_{ {\bf q}}\bar\Phi\right]
+\nabla_{{\bf q}} {p}   
&=&0 \\&&\\
3\bar\Phi''-\Delta_{{\bf q}}\bar\Phi
    -6{\mathcal H}^2\bar\Phi&=&{4\pi G a^4}\delta T, \end{array} \right.
    \end{eqnarray}
where we continue using the total derivative $D_{\eta}f=\partial_{\eta}f+
{\bf u}.\nabla_{{\bf q}}f$.

}

\

{And introducing the variable $\Psi\equiv a^2(1-2\Phi)$ the last equation can be written as
\begin{eqnarray}
  3\Psi''-\Delta_{{\bf q}}\Psi
    -6{\mathcal H}^2 \Psi&=&-{8\pi G a^4} T.  
\end{eqnarray}

Therefore, the Newtonian equations in an expanding universe can be written as:
\begin{eqnarray}
\label{newtonequations3}
\left\{ \begin{array}{ccc}
 D_{\eta}{\epsilon}+(\epsilon+p)\left[3\left({\mathcal H}+ \dfrac{1}{2}{\left(\dfrac{\Psi}{a^2}\right)'} \right)
+\nabla_{\bf q}.  {\bf u}\right]   &= &0 \\&&\\
D_{\eta} \left( {(\epsilon+p)}{{\bf u}}    \right)+
(\epsilon+p)\left[4{\mathcal H}{\bf u}
+ (\nabla_{{\bf q}}.{\bf u}){\bf u} -\dfrac{1}{2a^2}\nabla_{ {\bf q}}\Psi\right]
+\nabla_{{\bf q}} {p}   
&=&0 \\&&\\
3\Psi''-\Delta_{{\bf q}}\Psi
    -6{\mathcal H}^2\Psi&=&-{8\pi G a^4} T,\end{array} \right.
    \end{eqnarray}
and in a static one, they become:
\begin{eqnarray}
\label{newtonequations3}
\left\{ \begin{array}{ccc}
 D_{t}{\epsilon}+(\epsilon+p)\left[ 
\nabla.  {\bf u}
-3\dot{\Phi}\right]   &= &0 \\&&\\
D_{t} \left( {(\epsilon+p)}{{\bf u}}    \right)+
(\epsilon+p)\left[(\nabla.{\bf u}){\bf u} +\nabla{\Phi}\right]
+\nabla {p}   
&=&0 \\&&\\
3\ddot{\Phi}-\Delta{\Phi}
    &=&{4\pi G } T.\end{array} \right.
    \end{eqnarray}


}


\section{Conclusions}


In the current investigation
have derived  the relativistic Friedmann equations, originating from the foundations of Newtonian mechanics and the first law of thermodynamics. Employing a Lagrangian formulation, we have expanded our inquiry to encompass the Friedmann equations for both fluid and scalar field scenarios. Noteworthy is the profound connection between the matter Lagrangian and energy density for a fluid, as well as its association with pressure in a universe housing a scalar field, a relationship persisting in its Newtonian counterpart as evidenced in our study. Moreover, our research has showcased the adaptability of this formulation by exploring its application to open systems, thus broadening the horizons of our investigation.

This alternative methodology for deriving the Friedmann equations furnishes a robust framework for probing the universe's dynamics and offers profound insights into the intricate interplay between diverse physical properties. By amalgamating Newtonian mechanics and thermodynamics, we stand poised to unravel the processes governing the evolution of our cosmos.

Lastly, our endeavors have extended to the derivation of perturbation equations in the Newtonian gauge within the realm of General Relativity, originating from the classical trio of equations governing a perfect fluid. These encompass the continuity equation, ensuring the conservation of mass, the Euler equation, governing the conservation of momentum, and the Poisson equation, delineating the gravitational potential's relationship with the energy density distribution. Through this comprehensive approach, we aim to bridge classical mechanics with the profound insights of General Relativity, thereby advancing our understanding of the universe's dynamic evolution.

\

\begin{acknowledgments}
JdH is supported by the Spanish grant 
PID2021-123903NB-I00
funded by MCIN/AEI/10.13039/501100011033 and by ``ERDF A way of making Europe''. 
\end{acknowledgments}

\end{document}